\documentclass[12pt]{article}
\usepackage{amssymb,amsmath} % packages from American Mathematical Society
\usepackage{graphicx} %eps figures can be used instead
\usepackage{cite}

\title{Broadband optical cooling of molecular rotors from room temperature to the ground state}
\author{Chien-Yu~Lien$^1$, Christopher~M.~Seck$^1$, Yen-Wei~Lin$^1$, Jason~H.~V.~Nguyen$^{1\dagger}$ \\
David~A.~Tabor$^1$, \& Brian~C.~Odom$^{1\ast}$\\
\\
\normalsize{$^{1}$Department of Physics and Astronomy, Northwestern University,}\\
\normalsize{2145 Sheridan Rd., Evanston, IL 60208, USA}\\
\normalsize{$^\dagger$Current Address: Department of Physics and Astronomy, Rice University,}\\
\normalsize{Brockman Hall for Physics, 6100 Main St., Houston, TX 77005}\\
\\
\normalsize{$^\ast$Corresponding author; E-mail:  b-odom@northwestern.edu.}
}

\date{}

\def\XdS{X$^{2}\Sigma^{+}$}
\def\AdP{A$^{2}\Pi$}
\def\AdPl{A$^{2}\Pi_{1/2}$}
\newcommand{\ket}[1]{\mbox{$|#1\rangle$}}
\newcommand{\Ba}{Ba$^+$}
\newcommand{\Al}{Al$^+$}
\newcommand{\AlH}{AlH$^+$}

\newcommand{\us}{$\mu$s}

\newcommand{\cm}{cm$^{-1}$}

\begin{document}

\maketitle

\begin{abstract}
Laser cycling of resonances can remove entropy from a system via spontaneously emitted photons, with electronic resonances providing the fastest cooling timescales because of their rapid spontaneous relaxation. Although atoms are routinely laser cooled, even simple molecules pose two interrelated challenges for cooling: every populated rotational-vibrational state requires a different laser frequency, and electronic relaxation generally excites vibrations. Here, we cool trapped AlH$^+$ molecules to their ground rotational-vibrational quantum state using an electronically-exciting broadband laser to simultaneously drive cooling resonances from many different rotational levels.  Undesired vibrational excitation is avoided because of vibrational-electronic decoupling in AlH$^+$. We demonstrate rotational cooling on the 140(20)~ms timescale from room temperature to $3.8^{+0.9}_{-0.3}$ K, with the ground state population increasing from ${\sim}3$\% to $95.4^{+1.3}_{-2.1}$\%. This cooling technique could be applied to several other neutral and charged molecular species useful for quantum information processing, ultracold chemistry applications, and precision tests of fundamental symmetries.
\end{abstract}

\newpage
Laser cooling of atomic translational and internal energies has made possible a wide range of applications including creation of novel phases of matter\cite{bec_book_2012}, atom-wave accelerometry\cite{dickerson_multiaxis_2013}, and ultraprecise timekeeping \cite{bloom_optical_2014}.  Because of their additional internal structure, laser cooled molecules offer possibilities for extending the quantum toolkit in completely new directions. For instance, the large electric dipoles of trapped polar molecules could be used to coherently transfer information between molecules or to electronic circuits for quantum information processing\cite{demille_quantum_2002, andre_coherent_2006}, ultracold molecular dipoles could be oriented by external fields to control chemical reactions\cite{lemeshko_manipulation_2013}, and molecular structure can provide enhanced sensitivity in searches for breakdowns of the Standard Model of particle physics\cite{schiller_tests_2005, demille_using_2008, leanhardt_high-resolution_2011, baron_order_2014}. However, before the full potential of molecules can be harnessed, methods are needed to prepare and reset their rotational and vibrational quantum states, with laser control offering a promising option.

Translational cooling of molecules initially in a particular rotational-vibrational state has been achieved using narrowband lasers, with unwanted vibrational excitation minimized because of an unusual degree of electronic-vibrational decoupling in the selected species\cite{shuman_laser_2010,hummon_2d_2013,zhelyazkova_laser_2013}. Vibrational cooling of translationally cold molecules has been performed using broadband lasers\cite{viteau_optical_2008} and incoherent light  sources\cite{sofikitis_vibrational_2009} with spectral filtering removing light that would lead to vibrational heating, and also by using spectrally tuned multimode diode lasers\cite{wakim_luminorefrigeration:_2012}. Rotational cooling is particularly critical for many applications because of the relatively small energy of rotational excitations; for instance, while many species are already in their ground vibrational state at room temperature, several to many rotational states are populated for all molecules. Rotational cooling has previously only been accomplished using narrowband lasers optically pumping population from a small number of rotational states at one time. A static narrowband laser was used to transfer population from two excited rotational states to the ground state in a beam experiment\cite{baron_order_2014}.  In an experiment on molecules formed by photoassocation from an ultracold atomic gas, cooling from the few populated rotational states to a single state was achieved by pumping from each level serially using a swept narrowband laser, in conjunction with a broadband laser providing vibrational cooling\cite{manai_rovibrational_2012}. Finally, rotational cooling of trapped molecular ions from room temperature to the ground state was achieved on relatively slow timescales (seconds) by waiting for spontaneous and blackbody radiation-driven rotational transitions to redistribute population to levels pumped by narrowband lasers\cite{staanum_rotational_2010,schneider_all-optical_2010}.

Broad-spectrum rotational cooling has previously been proposed \cite{sofikitis_molecular_2009, nguyen_prospects_2011,nguyen_challenges_2011, lien_optical_2011}, and here we demonstrate its implementation, using a broadband laser tuned to an electronic resonance to cool rotations of trapped \AlH\ molecules from room temperature to their ground rotational state.  As in the previous molecular translational cooling demonstrations\cite{shuman_laser_2010,hummon_2d_2013,zhelyazkova_laser_2013}, undesired vibrational excitation during electronic-resonance cycling is avoided because of vibrational-electronic decoupling in our chosen molecule, AlH$^+$. Broadband rotational cooling will allow rapid quantum state preparation both in molecular beam experiments, in which the time available for laser cooling is limited, and in trapped-molecule experiments where the initial rotational temperature is too high for cooling by a swept narrowband laser. In non-destructive trapped-molecule work such as spectroscopy or quantum information processing, particularly in ion traps where hold times can extend to many hours, broadband rotational cooling could also provide rapid state resets allowing the same molecules to be reused repeatedly, rather than requiring the trap to be dumped and reloaded in each experimental iteration.

\section*{Results}
\subsection*{Choice of molecular species}
\AlH\ is a favorable species for broadband rotational optical cooling (BROC)\cite{lien_optical_2011} because its \AdP-\XdS\ electronic resonance is fairly decoupled from vibrations, supporting around 30 excitation/spontaneous emission cycles before vibrational excitation occurs\cite{nguyen_challenges_2011} (Fig.~1). It is also important to consider the complexity of the spectral filtering required to ensure that the broadband laser drives only rotational cooling transitions\cite{sofikitis_molecular_2009}. For \AlH\ and similar species, only simple spectral filtering is needed because vibrational-electronic decoupling implies matched rotational constants in the two electronic states, and thus well-separated rotationally de-exciting (P-branch) and de-exciting (R-branch) bands (Fig.~2).

\subsection*{Trapping and rotational analysis apparatus}
Our experiment is performed in a room temperature linear Paul trap integrated with equipment for time-of-flight mass spectrometry (TOFMS). In each experimental run, laser ablation is used to load laser cooled \Ba\ ions along with typically 50 \Al\ ions, with the \Al\ numbers kept small to maintain sufficient TOFMS resolution. We operate with an intentionally degraded vacuum ($3\times10^{-10}$ mbar) so that \AlH\ is formed on a one-minute timescale by reaction of \Al\ with background gas.  Coulomb repulsion keeps ions several microns apart, preventing ion-ion chemical reactions and allowing the laser-cooled \Ba\ ions to sympathetically cool \AlH\ ions into a so-called Coulomb crystal\cite{bowe_sympathetic_1999} with a translational temperature $< 1$ K.  The large inter-ion separation and ultrahigh vacuum conditions result in decoupling of the \AlH\ translational, rotational, and vibrational temperatures. Optical rotational cooling could also be performed on translationally warmer samples, but crystallization is advantageous because it localizes the molecules, allowing for higher laser intensity and improving TOFMS resolution.

After loading the trap, \AlH\ rotations and vibrations equilibrate to room temperature via blackbody radiation before cooling lasers are applied (Fig.~3). At room temperature, 99.9\% of the \AlH\ population is in the lowest vibrational state $v=0$, with significant population distributed among the first ten rotational levels, $N=0-9$, and 4\% in $N\ge10$. The rotational population distribution is destructively probed by state-selective ($1 + 1'$) resonance-enhanced multiphoton dissociation (REMPD)\cite{bertelsen_rotational_2006, seck_rotational_2014}, which converts \AlH\ only in a target rotational-vibrational state into \Al, and is analyzed by TOFMS.  In each experimental iteration the REMPD probe is tuned to a single rotational level, and \Al\ and \AlH\ populations are counted. We do not attempt to independently analyze the number of \AlH\ loaded into the trap, but rather determine the degree of rotational cooling by TOFMS measurements of the \Al\ to \AlH\ ratios after REMPD. The trap is reloaded and the same experiment repeated typically 10 times to gather statistics for each rotational level. Further details of the apparatus and state readout can be found in Ref.~\citen{seck_rotational_2014}.

\subsection*{Spectral filtering of the broadband laser}
The light source for BROC is a frequency-doubled femtosecond laser (SpectraPhysics Mai Tai) yielding 900 mW at the 360 nm A$^{2}\Pi_{1/2}$-\XdS\ transition. The laser is spectrally filtered after the doubling stage. Spectral filtering is accomplished using a home-built device, consisting of diffraction gratings and cylindrical lenses in the 4-f Fourier-transform optical layout \cite{weiner_femtosecond_2000} and a razor blade mask to block the high-frequency part of the spectrum. The broadband spectral filtering apparatus and its application to rotational cooling is described in detail in Ref.~\citen{lien_optical_2011}.

We achieve a spectral filtering cutoff sharper than observable in the commercial spectrometer data shown in Fig.~2.  Two independent characterizations described in Methods yield a cutoff resolution of $2\ \textrm{cm}^{-1}$ at the 10 dB extinction point. The ultimately achievable spectral resolution is set by the diffraction limit; for our grating (3600 lines/mm) and beam waist (16 mm), this limit is $0.1\ \textrm{cm}^{-1}$ at full width at half maximum\cite{lien_optical_2011}.  The dominant effect making our current resolution far from the diffraction limit is the large spherical aberration in our off-the-shelf lenses; commercially available optics are readily available to bring the cutoff resolution near the diffraction limit.  The spectral separation between rotationally heating and cooling transitions is approximately $2 B_\textrm{e}$ ($4 B_\textrm{e}$) for species with (without) allowed Q-branch (rotationally non-changing) transitions, where $B_\textrm{e}$ is the rotational constant for the molecule. For \AlH, $2 B_\textrm{e} = 13$ \cm, so even the currently achieved filtering resolution is sufficient for efficient cooling.

The spectrum of the femtosecond laser is actually a series of ``comb teeth" spaced by 80 MHz, so a mechanism is needed to guarantee that cooling transitions do not accidentally fall between the teeth.  In our case, ion translational heating occurs for the radially extended Coulomb crystal due to coupling to the radiofrequency trapping voltage\cite{wineland_experimental_1998}, yielding  sufficient Doppler broadening to ensure spectral overlap\cite{lien_optical_2011}.  Alternately, frequency modulation of the cooling laser could be used to modify its tooth structure, eliminating the need for Doppler broadening.

\subsection*{Rotational cooling schemes}
The angular momentum selection rules for the \AdPl-\XdS\ rotational cooling transition are $\Delta J = 0, \pm 1$, where $\textbf{J} = \textbf{N} + \textbf{S} + \textbf{L}$ is the total angular momentum including molecular rotation $\textbf{N}$, electron spin $\textbf{S}$, and electronic orbital angular momentum $\textbf{L}$. Spectral filtering removes all but rotationally de-exciting P-branch transitions (Fig.~2), driving rotational cooling. Each electronic excitation and spontaneous emission cycle conserves molecular parity, so in the simplest BROC spectral filtering configuration, population is cooled to the lowest rotational-vibrational state of each parity, \ket{\textrm{X}^2 \Sigma^+, v=0, N=0} and \ket{\textrm{X}^2 \Sigma^+, v=0, N=1} (Fig.~3a).  Because of imperfect electronic-vibration deoupling in \AlH, some undesirable vibrational excitation (14\% at the fast-timescale equilibrium) also occurs during this cooling process.  On the longer timescales (${\sim}100$ ms) of vibrational spontaneous emission, all population is pumped to the two lowest states of each parity with no residual vibrational excitation.

In order to cool all population into the positive-parity \ket{\textrm{X}^2 \Sigma^+, v=0, N=0} level, we employ a modified vibrationally-assisted broadband rotational optical cooling scheme (VA-BROC); here, parity conservation of BROC is destroyed by forcing a vibrationally-exciting electronic decay to occur only for the negative parity. The spectral cutoff frequency is shifted slightly (Fig.~2b) to additionally drive the \ket{\textrm{X}^2 \Sigma^+, v=0, N=1} $\rightarrow$ \ket{\textrm{A}^2 \Pi_{1/2},v'=0, (N')=0} transition, which has no alternate channel for spontaneous relaxation within the $v=0$ manifold, resulting in electronic cycling, until spontaneous electronic relaxation to $v=1$ and subsequent spontaneous vibrational relaxation flips the parity. Population immediately afterwards is entirely in the positive-parity \ket{\textrm{X}^2 \Sigma^+, v=0, N=0} or \ket{\textrm{X}^2 \Sigma^+, v=0, N=2} levels, with population in the latter quickly pumped to the ground state (Fig.~3a).  Note that only a single spectral filtering configuration is needed for VA-BROC cooling from room temperature to the ground state, since the previously discussed BROC process is included within VA-BROC.

Since the upper state degeneracy is smaller than that of the lower state when driving P-branch transitions, dark states must also be considered. The unpaired electron spin in the \XdS\ state allows modest magnetic fields to cause sufficient Larmor precession rates and prevent a slowdown of rotational cooling.  In our experiment, we use a ${\sim}2$ Gauss magnetic field perpendicular to the cooling laser polarization.

\subsection*{Rotational cooling results}
Fig.~3 shows the measured rotational distributions before and after 5~s of BROC or VA-BROC illumination. The initial rotational populations are well represented by a thermal distribution, given by a Boltzman exponential times a degeneracy factor, validating the analysis techniques.  A ``toy" Monte Carlo simulation described in Methods is used for statistical analysis.

BROC cools each parity independently and yields 48(4)\% population in \ket{\textrm{X}^2 \Sigma^+, v=0, N=0} and 46(3)\% in \ket{\textrm{X}^2 \Sigma^+, v=0, N=1}. These data are statistically consistent with fully efficient BROC cooling, as are REMPD probes finding no population in higher rotational states. Fig.~3 does include data from an anomalous run yielding non-zero $N=3$ population; however, as discussed in Methods, this signal is most likely from detector noise.  Because the spacing between consecutive states of the same parity is large, 50 K (90 K) between the lowest two states of positive (negative) parity, only relatively weak constraints can be set on the temperature achieved by BROC.  The 90\% CL temperature upper limits are 13 K for positive parity and 19 K for negative parity.  The negative-parity limit is obtained when the suspect $N=3$ signal is neglected; if that signal is attributed to imperfect BROC cooling, then the corresponding measured temperature for negative parity would be 19 K (where the numerical correspondence to the limit-value is coincidence).

After VA-BROC, the population measured in \ket{\textrm{X}^2 \Sigma^+, v=0, N=0} is $95.4^{+1.3}_{-2.1}\%$; probing \ket{\textrm{X}^2 \Sigma^+, v=0, N=1}, we measure a population of $2.0^{+3.3}_{-1.4}$\%, and we observe no population in $N=2$ or $N=3$.  These results are consistent with all population after VA-BROC being in the two lowest rotational states, with the population ratio corresponding to a rotational temperature of $3.8^{+0.9}_{-0.3}$~K.

Although simulation\cite{lien_optical_2011} predicts that the BROC phase of cooling occurs as fast as 8~\us\ with our current laser intensity, population dynamics measurements were limited by the 10 ms speed of a mechanical shutter gating the cooling light.  Nonetheless, this timing resolution allows us to analyze the slower of the two VA-BROC phases associated with a vibrational spontaneous relaxation event.  Fig.~3b (inset) shows data taken for various cooling times with the population measurement occurring 1 s (several vibrational-relaxation lifetimes) later. Similarly to the approach used in Ref.~\citen{versolato_decay_2013}, the vibrational lifetime $\tau$ can be extracted from the time-dependent population.  The fit described in Methods yields $\tau = 140(20)$ ms, in good agreement with the theory value of 127 ms\cite{nguyen_challenges_2011}. Blackbody-radiation induced rotational and vibrational transitions from low-lying rotational states occur on ${\sim}100$ s timescale, so these processes do not significantly affect cooling efficiency or analysis.

\section*{Discussion}
We expect that our rotational temperature after VA-BROC could in the future be reduced by improved spectral filtering. Regarding improvements in timescale, we note that BROC already cools to the lowest two rotational states in as fast as microseconds, potentially an acceptable starting point for some applications requiring cooling faster than the ${\sim}100$ ms VA timescale. Cooling to a single rotational state could also be achieved in a few microseconds by performing parity cooling without a vibrational relaxation event.  For instance, population pumped by BROC into \ket{\textrm{X}^2\Sigma^+,v=0,N=1} could be driven by a two-photon excitation to \AdP, with subsequent electronic spontaneous relaxation accomplishing a parity flip in less than a microsecond. An additional femtosecond laser providing vibrational cooling could reduce or effectively eliminate transient vibrational population buildup during BROC due to imperfect vibrational-electronic decoupling.  Unremediated vibrational excitations during BROC will be more of a concern in heavier reduced-mass species, both because of slower vibrational spontaneous relaxation rates and because of the larger number of rotational cooling steps needed to cool from a given temperature. Further details of these improvements are discussed in Methods.

It is instructive to compare the performance of BROC with a few other approaches for rotational cooling of trapped molecules.  The only previous laser cooling of rotations from room temperature used narrowband excitation of vibrational resonances, yielding ground state populations of 36.7\% (20~K) for MgH$^+$ (Ref.~\citen{staanum_rotational_2010}) and 78\% (27~K) for HD$^+$ (Ref.~\citen{schneider_all-optical_2010}); the process-limiting timescale of several seconds was associated with rotational transitions.  In comparison, VA-BROC cooling of \AlH  achieves 95\% ground state population and a temperature of 3.8 K, with a process-limiting timescale of ${\sim}100$ ms associated with a vibrational transition. Comparing generality of the approaches, the narrowband technique is readily applicable to any polar hydride but has limited utility for species with larger reduced mass (i.e. fluorides or chlorides) due to their slower vibrational and rotational transition rates\cite{schneider_all-optical_2010}.  BROC is most readily applicable to specialized molecules with decoupled vibrational and electronic modes, but they can be heavier reduced-mass species. Finally, compared with direct cryogenic buffer gas cooling of molecular ions in a 4 K apparatus\cite{hansen_efficient_2014}, we obtain in our room-temperature apparatus a temperature a few times colder for a single molecule (and many times colder for larger collections of molecules) on a similar ${\sim}100$~ms timescale.

Several other neutral and ionic candidate species for the simplest implementation of BROC have been identified\cite{rosa_laser-cooling_2004, nguyen_prospects_2011,nguyen_challenges_2011}.  It is interesting to consider the spectral filtering resolution required to rotationally cool diatomic species with heavier reduced mass and correspondingly smaller rotational constants and more congested spectra.  Working with the same beam size but near the diffraction limit, we could expect to achieve enough resolution for efficient cooling of fluorides to their ground state and partial cooling of chlorides.  Further resolution improvements could be obtained by using a finer grating and possibly by utilizing interference filters to handle transmission of the spectrum around the cutoff (D. Comparat, personal communication). Another option for rotational cooling of species with very small rotational constants would be to use a swept CW laser to cool population already near the ground state, as used in Ref.~\citen{manai_rovibrational_2012}, but adding to their technique a BROC laser to cool higher-lying population.

In analogy with the numerous applications arising from translational laser cooling of only a handful of atomic species, we envision many new possibilities coming from rotational laser cooling of select molecular species with vibrational-electronic decoupling. Additionally, the combination of BROC with broadband vibrational cooling\cite{viteau_optical_2008} could make possible optical rotational cooling of a still larger class of molecules. Building upon the current work, \AlH\ is a good candidate for many currently unrealized goals for trapped molecules, including state-to-state ultracold chemistry, all-optical cooling of all internal and external molecular degrees of freedom, single-molecule fluorescence detection, coherent transfer of quantum information between molecular rotations and external circuits\cite{andre_coherent_2006}, and single-molecule spectroscopy\cite{schmidt_spectroscopy_2005, lin_resonant_2013, wan_precision_2014} including laboratory searches for time-varying fundamental constants\cite{schiller_tests_2005}.

\section*{Methods}

\subsection*{Spectral filtering characterization}
Using a similar femtosecond laser and spectral filtering apparatus operating at a slightly different wavelength (385 nm) we use two different analysis techniques to more precisely characterize the spectral filtering resolution.  First, we create a home-built spectrometer by inserting a line-CCD camera at the Fourier plane. A continuous-wave laser is used to calibrate the spectrometer's frequency reading and resolution. Second, the narrowband laser is beat against the filtered broadband light on a photodetector. The lowest-frequency beat note occurs between the narrowband laser and the nearest comb tooth from the femtosecond laser.  From this beat note intensity, we determine the spectral density of the filtered broadband light at the narrowband reference laser's frequency. We then map the filtering behavior as we scan the cutoff across the reference. Both techniques yield a consistent result of $2\ \textrm{cm}^{-1}$ filtering cutoff sharpness at 10 dB extinction.

\subsection*{Anomalous point in BROC data set}
In one REMPD-TOFMS probe of $N=3$ population after BROC, for which there should be no \Al\ signal if BROC worked perfectly, we observed a structureless peak in the \Al\ bin of the TOFMS. The peak had a total charge consistent with a 2-ion \Al\ signal, which occur in our TOFMS data both with and without structure.  However, the peak was also consistent with structureless detector noise spikes observed occasionally at random times.  Given our overall sample size and typical TOFMS timing distribution, it is very unlikely that the only observed signal from a residual population in $N\ge2$ occurs as a single 2-ion event rather than as a small number of 1-ion events. Nonetheless, we take the conservative approach of treating this point as $N=3$ population resulting from imperfect BROC; we include it as such in Fig.~3b and in all analysis, except where otherwise noted.

\subsection*{Statistical analysis techniques and results}
A ``toy" Monte Carlo technique was used to perform statistical error analysis.  The method determines statistical properties by simulating repetition of a given  experiment a large number of times, each time with the same total \AlH\ number but with statistical fluctuations creating a distribution of simulated observations. An $n$-tuple array using binomial-distribution random number generation is created; each row in the array represents a set of independent measurements, e.g. $N=0$ and $N=1$ population fractions. Using variations of this analysis, experimental consistency with our model can be verified, and we can determine statistical bounds without relying on binomial confidence intervals far from the central limit approximation. Example simulation results are shown in Fig.~4.

Perfect BROC cooling would result in 100\% population in $N = 0$ and $N = 1$.  In our experiment, the combined population measured to be missing from these two states 6(5)\%, consistent with the expectation of fully efficient cooling and no loss from molecular dissociation\cite{nguyen_challenges_2011}.  Apart from the suspect $N=3$ data point discussed above, no REMPD probes after BROC detected population in $N>1$, which is also consistent with fully efficient BROC cooling.  Finally, to verify agreement of the BROC measurements with our rate-equation simulation \cite{lien_optical_2011}, an $n$-tuple was generated that uses the experimental total ion numbers for our $N = 0$ and $N = 1$ BROC data sets, along with the BROC population fractions of $f_0 = 48.3$\% in $N = 0$ and $f_1 = 51.7$\% in $N = 1$ predicted from simulation.  For each of 100,000 entries in the $n$-tuple, the population fraction ratio, defined as $f_0 / f_1$, is calculated and histogrammed (Fig.~4a).  For this sample size, the data are well described by a Gaussian distribution; the experimentally observed ratio is consistent with the simulation, differing by 1.3 $\sigma$.

The energetically allowed process of predissociation, in which \AlH\ in the \AdP\ state dissociates into \Al\ and H, is expected to be rare \cite{nguyen_challenges_2011}.  Predissociation during BROC would yield a signal of excess population in all rotational states, since it would provide an additional channel besides REMPD for conversion of \AlH\ to \Al.  To set an upper limit on the predissociation fraction, the BROC total ion number detected in $N \ge 2$ was used, discarding the data set with suspect $N=3$ signal discussed above.  Using five $n$-tuples of 100,000 entries, an additional non-zero predissociation population was added to the toy Monte Carlo simulation; this parameter was varied by hand until only 10\% of $n$-tuple entries resulted in the observed value of 0 predissociated ions, from which we conclude at the 90\% confidence level (CL) that less than 0.2\% of the population predissociated during BROC.  We estimate from the rate-equation simulation that there are on average 5 excitations to \AdP\ for BROC cooling from a room temperature distribution.  Combining these results, we conclude at the 90\% CL that the branching ratio of predissociation from \AdP\ relative to spontaneous emission is $< 0.04\%$, corresponding to a 90\% CL upper limit on the predissociating decay rate of $2\pi \times 800$ Hz.

A similar method was used to set the BROC temperature upper limit for each parity state, described in Results.  An input temperature was used to calculate the thermal probabilities for being in the first excited state of each parity.  This probability was then used to generate five 100,000-entry $n$-tuples for the experimental sample size.  The temperature was increased until 10\% of the $n$-tuple entries contained 1 molecule in the second-lowest rotational state.

To determine the limits on possible population loss to unprobed states during VA-BROC, we consider the statistics of the population successfully cooled to $N=0$, and we ask whether it is consistent with the observed population remaining in $N=1$.  We define a loss fraction $f_x$ as the population in any state other than $N=0$ and insert this loss term into the Monte Carlo simulation.  The 68\% CL limits on $f_x$ were found by adjusting $f_x$ until only 16\% of the simulated distribution fell on either side of the observed value of of 4.6\% missing from $N=0$; a sample $n$-tuple distribution at one of these limits is shown in Fig~4b. The resulting value for VA-BROC loss is $4.6^{+2.1}_{-1.3}$\%.  A separate analysis using a similar technique for the population in $N=1$ yields $2.0^{+3.3}_{-1.4}$\%.   We conclude that our results are consistent with all population after VA-BROC being in either $N=0$ or $N=1$, with no evidence for loss to other states.

The population ratio between $N = 1$ and $N = 0$ was used to determine the VA-BROC experimental temperature. The 68\% CL limits, determined similarly to the VA-BROC loss and using five $n$-tuples of $100,000$ entries, yields a VA-BROC temperature of $3.8^{+0.9}_{-0.3}$ K.  An alternative analysis technique would be to use the populations rather than the ratio to determine a temperature.  For example, using the measured ground state population of $95\%$ would yield a rotational temperature of 4.4 K.  However, since extraction of the temperature from the population ratio makes no assumptions about detection efficiency, we consider it to be the more reliable technique.

\subsection*{Efficiency of VA-BROC cooling}
Apart from the suspect data point discussed above, there is no other evidence that imperfect BROC performance leaves population in $N\ge2$.  However, we observe a clear signal of population not fully cooled to $N=0$ by VA-BROC (Fig.~3b). This type of asymmetry between BROC and VA-BROC performance is expected because the spectral cutoff of the latter is necessarily closer to the Q-branch heating transition frequencies.  A rate-equation simulation predicts that for operation below saturation on all transitions, if 1\% of the intensity pumping P(1) leaks through the filter to pump Q(0), the result will be 6\% population in $N=1$, similar to the measured quantities discussed below.

We have not attempted to make a serious model for spectral leakage, but a simple analysis suggests plausibility for the presence of percent-level leakage intensity at Q(0). The separation between the P(1) and Q(0) transitions is 13 cm$^{-1}$, which is smaller than the accuracy of our absolute spectral calibration. The 2 cm$^{-1}$ width of this cutoff is large enough compared with the P(1)-Q(0) separation that percent-level leakage at Q(0) for non-optimized mask placement would be unsurprising.  Further investigation is required to confirm expectations that improved frequency tuning and cutoff resolution of the filter will lead to lower rotational temperatures after VA-BROC.

\subsection*{VA-BROC cooling-time model}
The detected population $P_0^d$ in $N=0$ is expected to obey $P_0^d = P_0^f-P_1^i B_2 e^{-T/\tau}$, where $P_0^f$ is the measured asymptotic population in $N=0$, $P_1^i$ is the initial population in $N=1$ following the faster BROC phase of cooling, taken here from simulation, $B_2 \approx 2/3$ is the calculated branching fraction for spontaneous emission from \ket{\textrm{X}^2 \Sigma^+, v=1, N=1} to \ket{\textrm{X}^2 \Sigma^+, v=0, N=2}, 1/$\tau$ is the overall vibrational spontaneous relaxation rate from $v=1$, and $T$ is the duration of the VA-PROC pulse. Note that the $N=0$ population at the shortest times probed is expected to be elevated from the $\sim50\%$ from BROC, because VA-BROC quickly pumps population from \ket{v=0, N=1} into \ket{v=1, N=0}, which then relaxes to either \ket{v=0, N=0} or \ket{v=0, N=2} before analysis.

\subsection*{Upgrades for faster cooling}
Driving a single two-photon transition from \XdS\ to \AdP\ would allow parity to be cooled without needing to resort to the relatively slow vibrational spontaneous relaxation event used in VA-BROC.  Driving this transition in microseconds could be accomplished with a single commercially available pulsed laser, using two photons of the same wavelength, far from any resonance.  If continuous parity cooling is desired, CW or quasi-CW lasers near-resonant with an intermediate state would be required.  Using one of the first few vibrational levels of \XdS\ as an intermediate state, commercially available lasers could drive the transition on microsecond timescales even if detuned from resonance by of order 100 MHz.  Although the hyperfine structure of \AlH\ relevant for such narrowband excitation is not yet known, all hyperfine states should be addressable with power-broadening \cite{bressel_manipulation_2012}.

\subsection*{Reducing vibrational excitation during BROC}
The problem of imperfect electronic-vibrational decoupling leading to ${\sim}14\%$ transient population excited to the $v>0$ on the BROC timescale is discussed in Results, as is the importance of this issue for BROC on heavier reduced-mass species. To achieve near-unity efficiency cooling on the BROC timescale, a second femtosecond laser can be added to drive some number of $\Delta v = -1$ vibrational cooling transitions.  This laser need not be spectrally filtered for typical hydrides and fluorides, since vibrational heating and cooling transitions are generally separated by more than the femtosecond laser bandwidth.  Although this is not the case for \AlH, other BROC-candidate molecules such as SiO$^+$\cite{nguyen_prospects_2011} have structure permitting a single laser to simultaneously drive vibrational cooling transitions from several excited vibrational levels.  We note that including a vibrationally cooling femtosecond laser might also be desirable for heavier-reduced mass species in order to deal with their higher thermally excited vibrational population at room temperature.

\bibliographystyle{naturemag}

\section*{Acknowledgments}
We thank Michael Schmitt for generous help with uncertainty analysis and Michael Drewsen and Stephan Schiller for useful conversations.  This work was supported by AFOSR grant no. FA9550-13-1-0116, NSF grant nos. PHY-1309701 and 0801685, and the David and Lucile Packard Foundation grant no. 2009-34713.

\section*{Author Contributions}
C.-Y.L. and C.S. designed and carried out the experiment.  Y.-W.L., J.N., and D.T. aided with the spectral filtering characterization, and J.N. developed the rate-equation simulation code. B.O. conceived the study and participated in all stages.

\newpage
\begin{figure}[htbp!]
  \centering
  \includegraphics[width=5.55in]{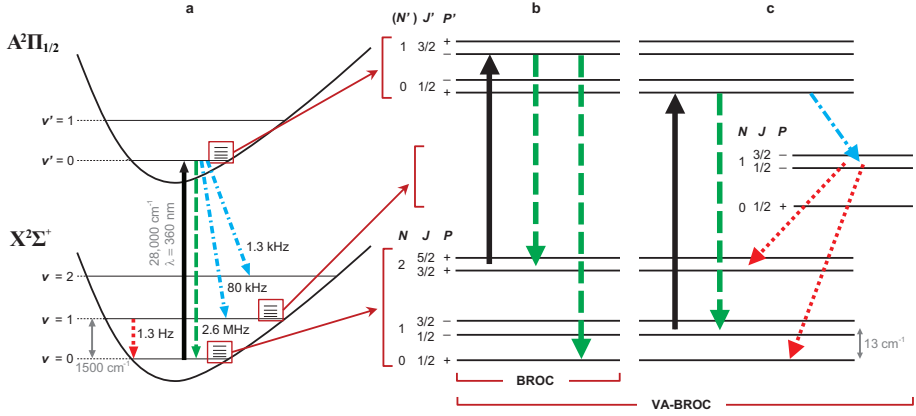}
  \caption{AlH$^+$ structure and broadband rotational cooling schemes. (\textbf{a}) The cooling laser drives electronic transitions from \mbox{$|\textrm{X}^2 \Sigma^+, v=0\rangle$} to \mbox{$|\textrm{A}^2\Pi_{1/2},v'=0\rangle$}(solid black upward arrows); electronic spontaneous relaxation most often occurs without vibrational excitation (dashed downward green arrows). Dash-dotted blue downward arrows indicate slower vibrationally-exciting electronic spontaneous relaxation, and dotted red downward arrows indicate much slower vibrational spontaneous relaxation. (\textbf{b}) BROC light achieves rotational cooling by exciting from the rotational state $N$ to $(N')=N-1$; although rotational angular momentum is not a good quantum number in A$^{2}\Pi$, $(N')$ serves as a convenient label. Parity and angular momentum selection rules allow for spontaneous relaxation only back to $N$ or to $N-2$.  Only excitation from the lowest state cooled by BROC is drawn in the figure. (\textbf{c}) The VA-BROC scheme additionally cools parity by including a spectral component to drive the first P-branch transition, driving electronic cycling until vibrational excitation results in a parity flip.}
  \label{broc}
\end{figure}

\begin{figure}[htbp!]
  \centering
  \includegraphics[width=3.75in]{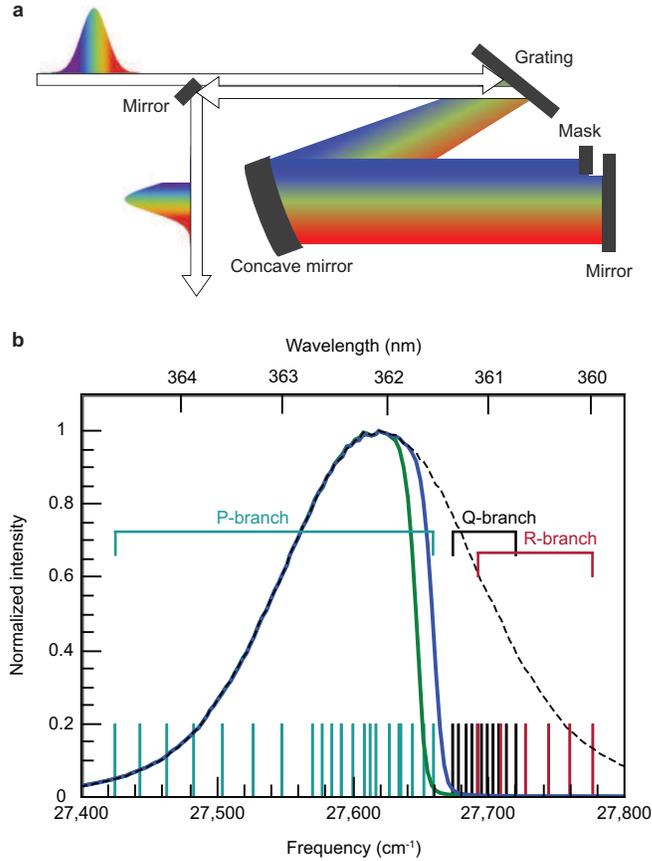}
  \caption{Spectral filtering of the broadband laser.
  (\textbf{a}) A mask at the Fourier plane of the so-called 4-f configuration commonly used in femtosecond pulse-shaping removes an undesirable portion of the broadband spectrum\cite{lien_optical_2011}.
  (\textbf{b}) Stick spectrum of the \AlH\ \AdPl-\XdS\ transition, with the unfiltered femtosecond laser (dashed), BROC (green), and VA-BROC (blue) spectra measured by a spectrometer (Ocean Optics HR4000).  The actual spectral cutoff resolution is sharper than the commercial spectrometer data shown in the figure. Since the spectrometer absolute calibration is not reliable to the required level, we use rotational cooling performance to fine-tune the mask position.}
  \label{spectrum}
\end{figure}

\begin{figure}[htbp!]
  \centering
  \includegraphics[width=3.5in]{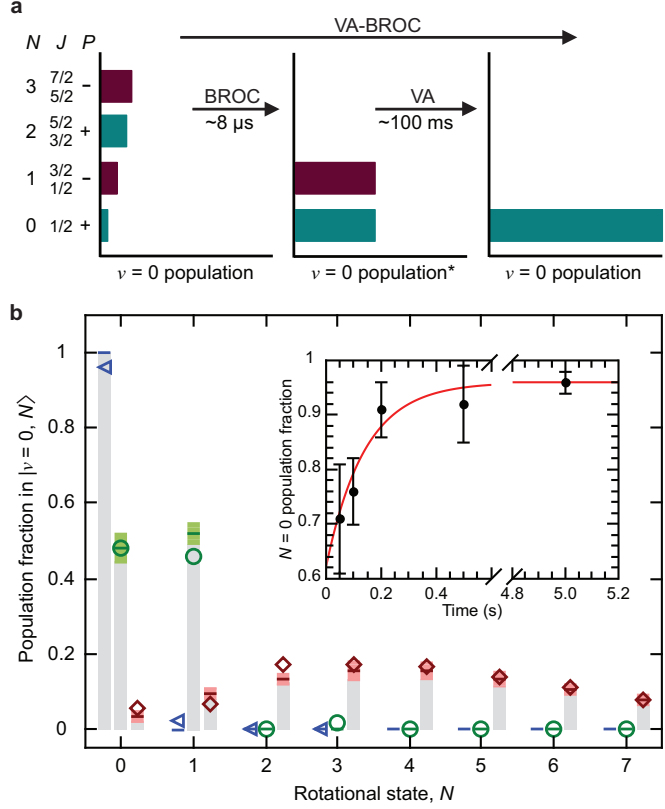}
  \caption{Rotational cooling results.
  (\textbf{a}) VA-BROC cooling contains two timescales: fast, parity-preserving electronic cycling and slower, vibrationally-assisted parity cooling. BROC excites a short-timescale $v>0$ population to an equilibrium value of 14\% (denoted by the $^*$), which then returns to ${\sim}0\%$ on the VA-BROC timescale.
  (\textbf{b})  Measured rotational populations for the initial thermal distribution (red diamonds), and after 5 s of cooling with spectral filtering in the BROC (green circles) and VA-BROC (blue triangles) configurations.  Expected thermal or simulated populations are indicated by horizontal lines with colored bands denoting the 68\% confidence regions for statistical uncertainties, for the experimental sample size, as determined by simulation.  Estimated systematic uncertainties are negligibly small in comparison.  Gray vertical bars are drawn to guide the eye.  Inset: Measured population fraction in $N=0$ versus duration of VA-BROC illumination; the single-parameter fit to the data is described in Methods. Vertical bars denote the $\pm 1 \sigma $ statistical confidence intervals.
  }
    \label{results}
\end{figure}

\begin{figure}[htbp!]
  \centering
  \includegraphics*[width=3.5in]{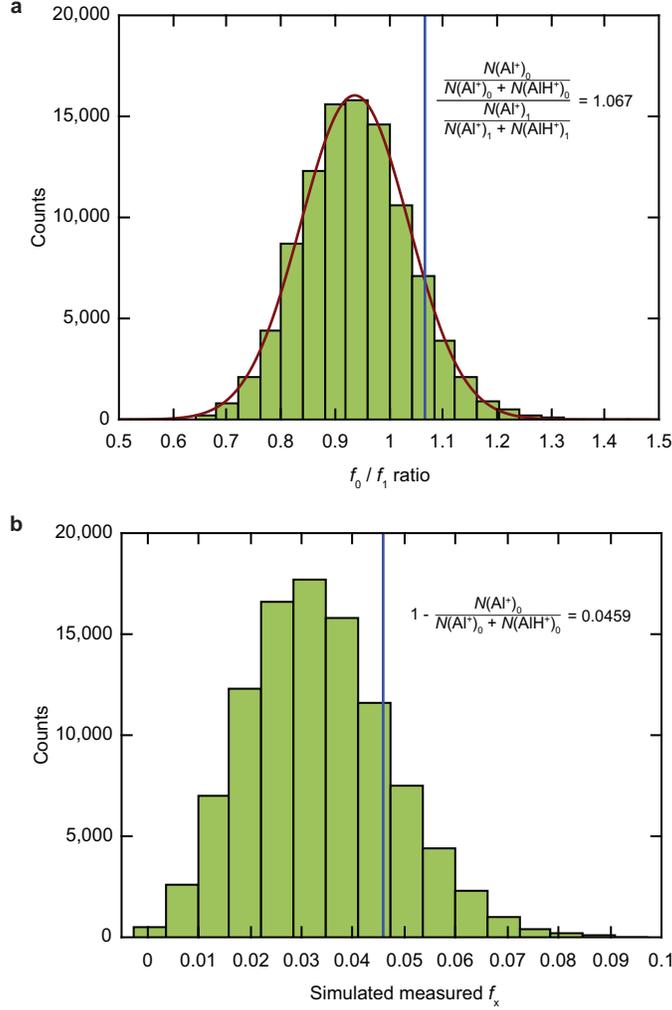}
  \caption{Example toy Monte Carlo simulation results.
  (\textbf{a}) Typical distribution of simulated ratio of the population fractions found in the two lowest states after BROC cooling. The expression for the ratio in terms of four observed ion counts, the observed population fraction (vertical line), and a Gaussian fit to the simulated distribution are also shown.
  (\textbf{b}) Typical histogram for determination of VA-BROC inefficiency $f_x$, defined as the fraction of population found to be missing from $N=0$.  The expression for $f_x$ in terms of two observed ion counts and the observed value (vertical line) are also shown. This histogram shows the Monte Carlo simulation results with the inserted inefficiency parameter turned down to find the lower bound of its 68\% confidence band, based on the measured $f_x$ and histogram areas.  The asymmetric distribution arises from small-number binomial statistics.}
  \label{monte}
\end{figure}

\end{document}